\documentstyle[aps,prl,tighten,floats,graphicx]{revtex}

\begin{document}

\twocolumn[
\hsize\textwidth\columnwidth\hsize\csname@twocolumnfalse\endcsname

\draft

\title{Two-scale analysis of the $SU(N)$ Kondo Model}
\author{D.~Villani, E.~Lange, A.~Avella\cite{Leave} and G.~Kotliar}
\address{Serin Physics Laboratory, Rutgers University, Piscataway, New Jersey 08855-0849, USA}
\date{December 20, 1999}

\maketitle

\begin{abstract}
We show how to resolve coherent low-energy features embedded in a broad
high-energy background by use of a fully self-consistent calculation
for composite particle operators. The method generalizes the
formulation of Roth, which linearizes the dynamics of composite
operators at any energy scale. Self-consistent equations are derived
and analyzed in the case of the single-impurity $SU(N)$ Kondo model.
\end{abstract}
]

\narrowtext

The development of effective methods for describing correlated electron
systems has been the subject of intensive activity over the last
decade, spurred by the experimental discoveries of the heavy fermion
systems, the high-temperature superconductivity and, generally, a
revival of interest in transition metal-oxide physics \cite{Imada:98}.

The Roth method for the correlation problem \cite{Roth:69} in the
context of the Hubbard model \cite{Hubbard:63} is based on an ansatz
which reduces the dynamics of field operators to a linearized one. The
essential idea is to select a basis of fermionic operators $\psi_i$,
write their equations of motion which involve operators $J_i$, and then
close these equations by projecting $J_i$ onto the basis by using the
Roth projector ${\cal P}$ defined by
\begin{equation}
{\cal P}\left(J_l\right)
=\sum_{rs}\left\langle\left\{J_l,\,\psi_r^\dagger\right\}\right\rangle
I^{-1}_{rs}\,\psi_s
\end{equation}
where
$I_{rs}=\left\langle\left\{\psi_r,\,\psi_s^\dagger\right\}\right\rangle$,
with $\left\{\_\,,\_\right\}$ denoting the anticommutator. In this
approach the determination of the Green's functions is then reduced to
the evaluation of certain static thermal averages:
$\left\langle\left\{\psi_r,\,\psi_s^\dagger\right\}\right\rangle$ and
$\left\langle\left\{J_l,\,\psi_r^\dagger\right\}\right\rangle$. When
these parameters are connected to matrix elements of the Green's
functions associated to the basis, one has a self-consistent scheme for
their calculation. However, this is often not the case, and further
approximations are introduced for their evaluation. The application of
this method, as well as similar methods \cite{Olds}, has recently been
reviewed by Mancini and collaborators \cite{Avella:98}. Through careful
comparison with existing numerical data, they concluded that good
results for many physical quantities are obtained by requiring that the
Green's functions fullfil exact equal-time identities accompanying the
fermionic character of the operators.

In spite of its intuitive appeal, there are several serious
difficulties with the Roth's method. Recent advances in the study of
correlated electron systems converge upon a picture of the one-particle
Green's function made up of incoherent broad spectral features in
addition to more dispersive quasi-particle bands which exist at lower
energies \cite{Bickers:87,Georges:96}. The Roth approach describes the
Green's functions in terms of a finite number of sharp poles which are
a poor description of the incoherent structure of the high-energy
spectra. Also, the presence of low-energy features embedded in a broad
high-energy background precludes the straightforward extension to
low-energy scales of this approach. Indeed, low-energy features cannot
be resolved increasing the size of the basis. Increasing the size of
the basis only amounts to calculating self-consistently a larger number
of spectral moments \cite{Mancini:98} which are dominated by
high-energy contributions. A clear example of this dramatic failure is
provided by the Kondo impurity model, where it has been proved
impossible to derive the existence of a Kondo resonance in the spectra
within a projection scheme.

In this Letter, we present a generalization of Roth's projection
technique which overcomes the limitations discussed above and, as an
illustration of the technique, we investigate the single-impurity
$SU(N)$ Kondo model \cite{Hewson:97}. Our goal is to introduce a
general technique in a simple context which is well understood, but so
far has not been successfully treated by techniques based on the
equations of motion. We will demonstrate that our method reproduces all
the well-known spectral features of the impurity model.

The method carries out the following steps. (i) In the first step, we
write the equations of motion for the operators of physical interest in
terms of higher order ones (or \emph{composite operators}). Similarly,
we express the Green's functions of interest in terms of the Green's
functions of the composite operators. The composite operators should
not have components on the physical fields at high energies. (ii) Then,
we evaluate the Green's functions of the composite operators by a
technique which is valid at high energies, such as the mode-coupling
approximation \cite{Bosse:78}. Use of the mode-coupling approximation
is motivated by the fact that it more clearly reflects that the
high-energy part of the spectra is quite incoherent. At this stage, we
divide the composite operators into (a) a high-energy part, described
by the mode-coupling approximation, and (b) a low-energy part, to be
determined by a non-perturbative closure of the equations of motion.
The necessity for this step is checked by writing the expressions for
the moments and noting that the mode-coupling approximation fails to
give the spectral weights, as expected from an independent evaluation
of the moments. (iii) The low-energy closure of the equations of motion
is dictated by the physics of the problem and is inspired by the
successes of the slave boson techniques \cite{Barnes:76}. It is a
simple quasi-particle theory which involves unknown parameters such as
the low-energy spectral weights. The self-consistent determination of
the low-energy parameters completes the full determination of the
physical Green's functions.

The $SU(N)$ Kondo model is described by the following Hamiltonian:
\begin{equation}
H=\sum_{{\bf k},{\bf k}'}c^\dagger({\bf k})\cdot\left[\delta_{{\bf
k}{\bf k}'}\,\varepsilon_c({\bf k})+2J_{\rm
K}\frac1{N\,N_s}\vec{\tau}_N\,\vec{n}^d\right]c({\bf k}')\label{E1}
\end{equation}
where $c({\bf k})$ denotes the conduction electron operator, and
$\vec{n}^d$ represents the spin operator at the impurity site
($\vec{n}^d\cdot\vec{n}^d=N(N+1)/2$). $\varepsilon_c({\bf k})$ and
$J_{\rm K}$ are the conduction electron energy and the Kondo coupling,
respectively. $N_s$ is the number of atomic sites of the host metal
responsible for the orbitals which form the conduction band. $\tau^a_N$
are the $N^2-1$ traceless generators of $SU(N)$
($\vec{\tau}_N\cdot\vec{\tau}_N=2(N^2-1)/N$). From (\ref{E1}), we have
\begin{equation}\label{E2}
{\rm i}\frac\partial{\partial t}c({\bf k})=\varepsilon_c({\bf
k})\,c({\bf k})+2J_{\rm K}\frac1{N\,N_s}\sum_{\bf
q}\vec{\tau}_N\cdot\vec{n}^d\,c({\bf q})
\end{equation}

Next, we introduce the Composite Heisenberg Field Operator:
\begin{equation}\label{E3}
\psi^\dagger=\left(\psi^\dagger_1\,, \psi^\dagger_2\right)
=\left(c^\dagger_0\,, 2\frac1Nc^\dagger_0\,
\vec{\tau}_N\cdot\vec{n}^d\right)
\end{equation}
where $c_0=\frac1{\sqrt{N_s}}\sum_{\bf q}c({\bf q})$ is the electron at
the impurity site. The field $\psi_2$ in (\ref{E3}) abides by the
criterion required by the method in (i). In fact, when
$\left\langle\left\{\psi_2,\psi_1^\dagger\right\}\right\rangle$ is
regarded as the scalar product of the field $\psi_2$ with the field
$c_0$, there is no component of $\psi_2$ on $c_0$ at any energy scale
since
$\left\langle\left\{\psi_2,\psi_1^\dagger\right\}\right\rangle=0$.
Then, using the equation (\ref{E2}) we can express the Green's
functions of the first field in terms of the Green's function for the
composite operator $\psi_2$. We have
\begin{eqnarray}
G_{11}(\omega)&=&\Gamma_0(\omega)+J_{\rm
K}^2\,\Gamma_0(\omega)\,G_{22}(\omega)\,\Gamma_0(\omega)\label{E4}\\
G_{12}(\omega)&=&J_{\rm K}\,\Gamma_0(\omega)\,G_{22}(\omega)\label{E5}
\end{eqnarray}
where $G_{\alpha\beta}(\omega)$ is the thermal Green's function
associated with the basis in (\ref{E3}) and
$\Gamma_0(\omega)=\frac1{N_s}\sum_{\bf
q}\frac1{\omega-\varepsilon_c({\bf q})}$ is the free propagator of the
field $c_0$. The total spectral weight attached to the second composite
field $\psi_2$ is
\begin{equation}\label{E9}
I_{22}=4\frac{N+1}{N^2}+4K_D
\end{equation}
where the Kondo amplitude
$K_D=\left\langle\psi_1\,\psi_2^\dagger\right\rangle=-2\frac1{N^2}
\left\langle\psi_1^\dagger\cdot\vec{\tau}_N\,\psi_1\,\vec{n}^d\right\rangle$
describes the binding between the localized spin and the spin
excitations of the field $c_0$. In order to resolve low-energy features
embedded in a high-energy background, we write
\begin{equation}\label{E6}
G_{22}(\omega)=G^H_{22}(\omega)+G^L_{22}(\omega)
\end{equation}
where $G^H_{22}(\omega)$ keeps the information about the band structure
and is not sensitive to features which are small with respect to the
bandwidth $2D$. In contrast, $G^L_{22}(\omega)$ mostly takes coherent
contribution from low energies and depends only weakly on the
high-energy part of the spectrum. Such a decomposition corresponds to
decomposition of the composite field $\psi_2$ as
$\psi_2=\psi^H_2+\psi^L_2$, with $\psi^H_2$ giving rise to incoherent
broad features, whereas $\psi^L_2$ emerges as an observable
quasi-particle at low energies.

In the high-energy regime, time-dependent correlation functions can be
treated within the mode-coupling approximation in terms of
electron-hole and charge-spin fluctuations. By use of mode-coupling in
the paramagnetic case, we have for the time ordered Green's function
$S^H_{22}(\omega)$:
\begin{equation}\label{E7}
S^H_{22}(\omega)=8\frac{N^2-1}{N^3}\frac{\rm
i}{2\pi}\int\!\!d\Omega\,S_{11r}(\omega-\Omega)\,S_{11}(\Omega)
\end{equation}
where
\begin{equation}\label{E8}
S_{11r}(t_i,t_j)=\left\langle{\cal
T}\left[n^d_r(t_i)\,n^d_r(t_j)\right]\right\rangle\ r=1,\ldots,N^2-1
\end{equation}
The spectral weight absorbed by the propagator $G_{22}(\omega)$ in the
mode-coupling form (\ref{E7}) is $4(N+1)/N^2$. For simplicity, we take
the atomic limit for the Bose propagator, so that
$G^H_{22}(\omega)=4\frac{N+1}{N^2}G_{11}(\omega)$. Other treatments for
the Bose propagators would not substantially affect our results for the
fermionic spectral function.

From the Hamiltonian (\ref{E1}) it is direct to derive
\begin{eqnarray}
{\rm i}\frac\partial{\partial
t}\psi_2=2\frac1N\vec{\tau}_N\cdot\vec{n}^d\,c_\varepsilon+2J_{\rm
K}\frac1N\vec{\tau}_N\cdot\vec{n}^d\,\psi_2\nonumber\\ +8{\rm
i}\,f_{abc}^N\,J_{\rm
K}\frac1{N^2}\tau_N^a\left[c_0^\dagger\cdot\tau_N^b\,c_0\right]n_c^d\,c_0\label{E10bis}
\end{eqnarray}
where $f_{abc}^N$ are the structure constants of the $SU(N)$ Lie
algebra ($\left[\tau_N^a,\tau_N^b\right]=2{\rm
i}\,f_{abc}^N\,\tau_N^c$) and
$c_\varepsilon=\frac1{\sqrt{N_s}}\sum_{\bf q}\varepsilon_c({\bf
q})\,c({\bf q})$. It is worth noting that the source (\ref{E10bis}) has
a direct component on the field $c_0$ ($2J_{\rm
K}/N\vec{\tau}_N\cdot\vec{n}^d\,\psi_2=4J_{\rm K}(N+1)/N^2 c_0+\ldots$)
which disappears for $N\rightarrow\infty$ being the coefficient
$4J_{\rm K}(N+1)/N^2$ (for $N=2$, $2J_{\rm
K}/N\vec{\tau}_N\cdot\vec{n}^d\,\psi_2\rightarrow 3J_{\rm K}c_0-2J_{\rm
K}\psi_2$).

In the low-energy regime, we assume the following dynamics for the
field $\psi_2$:
\begin{equation}\label{E10}
{\rm i}\frac\partial{\partial t}\psi^L_2=\aleph\,\psi_1
\end{equation}
This corresponds to the physical assumption that at low energies (i.e.,
at energies much smaller than $J_{\rm K}$ and $D$) we have a
quasi-particle theory. Indeed, the ansatz in (\ref{E10}) can be
described as an application of Roth's projection idea to a field
$\psi^L_2$ which has most of its spectral weight at low energies. Thus,
the high-energy spectral weight is already accounted for by the
mode-coupling approximation. This is the second main departure from the
original Roth approach, where an equation of motion for the field
$\psi_2$ would have been projected onto $\psi_1$ and $\psi_2$ itself.
This field would have spectral weight at all frequencies (e.g.,
$4(N+1)/N^2$) and from it a Kondo scale cannot be estimated. Once
again, noteworthy is the fact that the basis defined in (\ref{E3})
alone is inadequate to capture both low- and high- energy physics of
the Kondo model once a Roth truncation is realized. At this level of
approximation, in the scattering matrix (i.e., $J_{\rm
K}^2G_{22}(\omega)$ ) there is only one energy scale that cannot mimic
a crossover between the two regimes. This is set by $I_{22}$ where the
high-energy spectral weight (i.e., $4(N+1)/N^2$) prevents the
low-energy scale from emerging.

By combining (\ref{E2}) and (\ref{E10}) it is direct to show that
\begin{equation}\label{E11}
G^L_{22}(\omega)=\frac{I^L_{22}}{\omega-J_{\rm
K}^2\,I^L_{22}\,\Gamma_0(\omega)}
\end{equation}
being $\aleph=J_{\rm K}\,I^L_{22}$, after projecting (\ref{E10}) on the
field $\psi_1$. We have defined $I^L_{22}$ as the spectral weight of
$G_{22}(\omega)$ in the low-energy region. Also, it is implicitly
assumed that
$\left\langle\left\{\psi^L_2,\psi^{H\dagger}_2\right\}\right\rangle=0$
because they span different energy sectors of the Hilbert space. In
conclusion, we have
\begin{eqnarray}
G_{11}(\omega)&=&\frac{\Gamma_0(\omega)}{1-4\frac{N+1}{N^2}J_{\rm
K}^2\,\Gamma_0^2(\omega)}\nonumber\\ &+&\frac{J_{\rm
K}^2\,\Gamma^2_0(\omega)}{1-4\frac{N+1}{N^2 }J_{\rm
K}^2\,\Gamma_0^2(\omega)}\frac{I^L_{22}}{\omega-J_{\rm
K}^2\,I^L_{22}\,\Gamma_0(\omega)}\label{E12}\\
G_{12}(\omega)&=&\frac{4\frac{N+1}{N^2}J_{\rm
K}\,\Gamma_0^2(\omega)}{1-4\frac{N+1}{N^2}J_{\rm
K}^2\,\Gamma_0^2(\omega)}\nonumber\\ &+&\frac{J_{\rm
K}\,\Gamma_0(\omega)}{1-4\frac{N+1}{N^2}J_{\rm
K}^2\,\Gamma_0^2(\omega)}\frac{I^L_{22}}{\omega-J_{\rm
K}^2\,I^L_{22}\,\Gamma_0(\omega)}\label{13}\\
G_{22}(\omega)&=&\frac{4\frac{N+1}{N^2}\Gamma_0(\omega)}{1-4\frac{N+1}{N^2}J_{\rm
K}^2\,\Gamma_0^2(\omega)}\nonumber\\ &+&\frac1{1-4\frac{N+1}{N^2}J_{\rm
K}^2\,\Gamma_0^2(\omega)}\frac{I^L_{22}}{\omega-J_{\rm
K}^2\,I^L_{22}\,\Gamma_0(\omega)}\label{14}
\end{eqnarray}
At this stage of the method, once $I^L_{22}$ is evaluated, the problem
is solved, as in point (iii) referred to above. From Eq.~(\ref{E9}) it
is clear that $I^L_{22}$ is connected to $K_D$. However, it is crucial
to note that it represents the low-energy spectral weight and is a
distinct object in respect to the $K_D$ parameter which also contains
contributions from energies of the order $J_{\rm K}$. For the
determination of this low-energy scale, we need to call upon some
self-consistent condition. Evaluating the Kondo amplitude $K_D$ using
$G_{12}(\omega)$
\begin{equation}\label{E16}
K_D=-T\sum_nG_{12}(i\,\omega_n)\ \ \ \ \omega_n=(2n+1)\pi\,T
\end{equation}
and inserting this into Eq.~(\ref{E9}), we give a relation between
$I_{22}$ and $I^L_{22}$ of the form $I_{22}=F\left[I_{22}^L\right]$. To
get the self-consistent equation, we estimate
$I_{22}^H=F\left[I_{22}^L=0,T=J_{\rm K}\right]$ which results in an
equation for the low-energy spectral weight of the form
$I_{22}^L=I_{22}-I_{22}^H$. In other words, we choose $J_{\rm K}$ as
the energy above which the one-particle Green's function is made up of
incoherent high-energy contributions with no relevant temperature
dependence. While the equations resemble the slave boson equations,
they differ in a significant way from them. These equations do not
introduce additional redundant phases and Lagrange multipliers, and
avoid all the difficulties associated with the treatment of the gauge
fields. The slave boson method has been very successful in obtaining
low-energy information. High-energy information can also be obtained by
performing fluctuations around the mean-field solutions, but this gets
increasingly difficult particularly in lattice models
\cite{Houghton:88}.

\begin{figure}[tb]
\begin{center}
\includegraphics*[width=8cm]{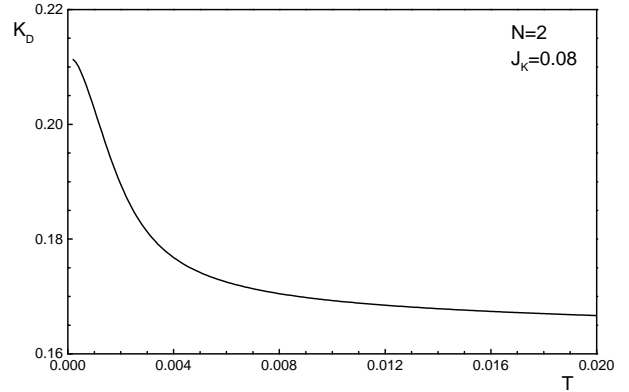}
\end{center}
\caption{The Kondo amplitude $K_D$ is plotted as a function of the
temperature for $J_{\rm K}=0.08$ and $N=2$.} \label{Fig1}
\end{figure}

\begin{figure}[tb]
\begin{center}
\includegraphics*[width=8cm]{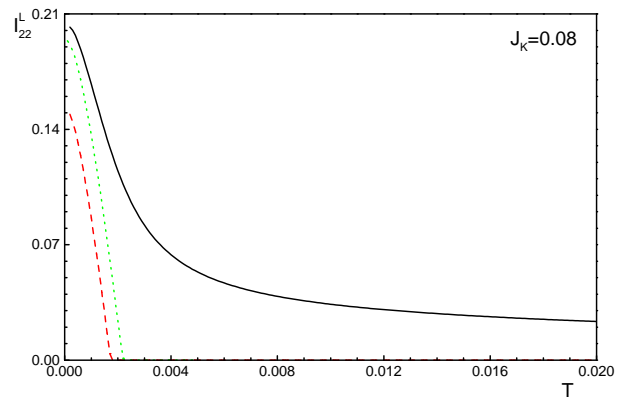}
\end{center}
\caption{The low-energy spectral weight $I^L_{22}$ is given as a
function of the temperature for $J_{\rm K}=0.08$ and $N=2$ (solid
line). As a point of reference, we give the solution after
$I_{22}^H=F\left[I_{22}^L=0\right]$ (dashed line) and for $N=\infty$
(dotted line).} \label{Fig2}
\end{figure}

\begin{figure}[tb]
\begin{center}
\includegraphics*[width=8cm]{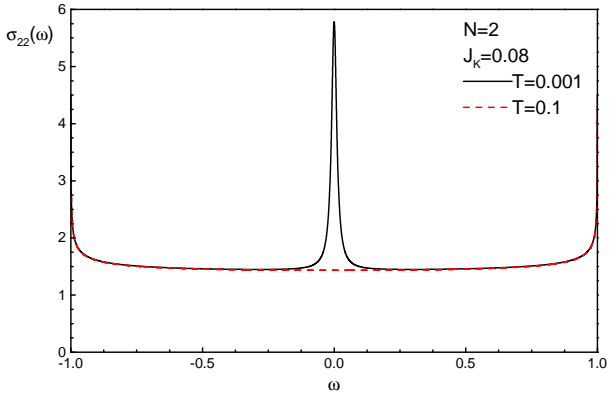}
\end{center}
\caption{The spectral density $\sigma_{22}(\omega)$ is reported for
$T=0.001$ (solid line) and $T=0.1$ (dashed line). $J_{\rm K}=0.08$ and
$N=2$.} \label{Plot3}
\end{figure}

We now present some results. For the numerical solution of the
equations we used a constant density of states
$\rho=\frac1{2D}\theta\left(D-|\omega|\right)$ for the field $c_0$ so
that $\Gamma^R_0(\omega)=\frac1{2D}\ln|\frac{D+\omega}{D-\omega}|-{\rm
i}\,\pi\frac1{2D}\theta\left(D-|\omega|\right)$. In Figs.~\ref{Fig1}
and \ref{Fig2}, the Kondo amplitude $K_D$ and $I^L_{22}$ are shown as
functions of the temperature for $J_{\rm K}=0.08$ and $N=2$. $D$ has
been set equal to $1$. In Fig.~\ref{Fig2} we also show the solution for
$N=\infty$ and the one after replacing
$I_{22}^H=F\left[I_{22}^L=0,T=J_{\rm K}\right]$ with
$I_{22}^H=F\left[I_{22}^L=0\right]$. Both these solutions give a
spurious transition at a characteristic temperature which is the
signature of the Kondo crossover in the exact solution. As in the slave
boson approximation, this is due to the absence of a small
inhomogeneous term which is present in our method and mimics mixing
effects between high- and low- energy contributions. The quantities in
Fig. \ref{Fig2} coincide only in the limit of large spin degeneracy
(i.e., $N\rightarrow\infty$) where the method recovers the exact
results \cite{Hewson:97}. Our numerical estimate for the Kondo
temperature $T_{\rm K}$ agrees well with the exact solution
\cite{Hewson:97,Andrei:83}. At this temperature ($T_{\rm
K}\simeq0.002$), $I_{22}^L$ has a change in the concavity of its slope
when plotted as a function of the temperature. In Fig.~\ref{Plot3}, we
present the spectral density
$\sigma_{22}(\omega)=\left(-\frac1\pi\right)\Im\left[
G^R_{22}(\omega)\right]$ for two different temperatures. Again, it is
clear that at high temperatures (i.e., $T\geq J_{\rm K}$) only a
high-energy incoherent background is left. A well-defined singlet
excitation mode no longer exists, even if some residual spin-spin
interaction persists being energetically favored by a finite bandwidth
(i.e., $K_D$ is non-zero at any temperature as in Fig.~\ref{Fig1}).

In conclusion, we have shown how to resolve coherent low-energy
features embedded in a broad high-energy background by use of a fully
self-consistent calculation for composite particle operators. In a
problem with more than one energy scale, which is typical of strongly
correlated systems, we succeeded to capture low-energy features. Our
scheme extends and improves upon the Roth's method by combining the
advantages of the methods based on the equations of motion and the
slave boson techniques. Finally, we note that when there is an
expansion parameter such as the size of the group, or the size of the
representation, our approach can be formulated so as to reduce to the
correct solution in the exactly soluble limit. We have illustrated this
here with the Kondo model in the limit of large spin symmetry group
which has often been shown to retain many crucial aspects of low-energy
physics \cite{Bickers:87}.

Finally, our approach is directly applicable to lattice models and work
in this direction is currently in progress.

\acknowledgments This work was supported by the NSF under Grant
DMR-95-29138. D.V. thanks Gina Valeri for her careful reading of the
manuscript.

\end{document}